# To the methodics of the superconducting nanotube paraconductivity calculation

Iogann Tolbatov

*Physics and Engineering Department, Kuban State University, Krasnodar, Russia*
(talbot1038@mail.ru)

*In the investigation carried out, the superconducting nanotube paraconductivity calculation algorithm is defined more precisely.*

The newest nanotechnologies are the fundament of the scientific-technological revolution in the XXI century. They can be developed to the mighty instrument of the Russia's technological complex integration in the international high technologies market. This state development direction is fixed in the governmental programme till the year of 2010.

Carbon nanotubes are the extended cylindrical structures with diameters ranging from one to several tens of nanometers and lengths up to several centimeters. They consist of one or more rolls of hexagonal planes of graphite (graphene) and end with a hemispherical head.

Unlike conventional three-dimensional conductors, a nanotube has a number of features [1] in such parameters as the effective dimensionality, electron-electron interaction, the degree of disorder, which can be explained by the fact that the fluctuation in the transport of univalent nanotubes is ballistic one-dimensional [2], id est, it can be described by the model of one-dimensional system of interacting electrons, known as Luttinger liquid [3].

Multivalent nanotubes consisting of several concentric graphite shells, exhibit properties consistent with diffusive transport - effects of weak localization in magnetoconductivity and the anomaly in tunneling density of states at zero voltage at the contact [4]. Similar phenomena were also found in bundles of univalent nanotubes. The superconducting state in nanotubes was found too [5,6].

Fluctuation transport in the superconducting nanotube means the presence of fluctuating Cooper pairs at temperatures above the superconducting transition temperature. The phenomenon of the fluctuation of transport is the cause of paraconductivity. Paraconductivity means singular contribution to the conductivity near the critical temperature. Calculation of paraconductivity in nanotubes has practical significance, because it allows to define their characteristics: thermopower, thermal conductivity at the threshold of transition to the superconducting state, the Hall effect.

Tensor for the fluctuation conductivity can be written in the following form [7]:

$$\sigma^{\alpha\alpha}(\varepsilon,H,\omega) = \frac{\pi}{2}\alpha e^2 T \sum_{\{i,l\}=0}^{\infty} R\left[\frac{\hat{v}_{\{il\}}^{\alpha} \cdot \hat{v}_{\{li\}}^{\alpha}}{E_{\{i\}}E_{\{l\}}\left(E_{\{i\}}+E_{\{l\}}-i\gamma_{GL}\omega\right)}\right]. \quad (1)$$

We assume that the reduced temperature and the reduced magnetic field respectively are the following: $\varepsilon \ll 1$, $h \ll 1$.

If we follow the methodology developed by Livanov and Varlamov [8], it should be in (1) at first the summation over index $\{i\}$ of angular quantization levels until the last (number $N$) filled level, and afterwards the integration of the result over the momentum along the axis of the cylinder.

If we use the values of the velocity matrix elements $\hat{v}_{\{pp'\}} = v_p \delta_{pp'}$, $\hat{v}_p = \frac{\partial E_p}{\partial p} = 2\alpha T_C \xi^2 p$, then we obtain as a result the general formula (1) transformed into the expression for the longitudinal component of nanotube paraconductivity:

$$\sigma^{||}(\varepsilon,H) = \frac{\pi\alpha e^2}{2S}T\int\frac{dp_{||}}{2\pi}\int\frac{dq_{||}}{2\pi}\sum_{i,l=-N}^{N}\frac{\hat{v}_{il,pq}^{||} \cdot \hat{v}_{li,qp}^{||}}{E_i(p_{||})E_l(q_{||})\left[E_i(p_{||})+E_l(q_{||})\right]} =$$

$$= \frac{\pi e^2}{16S}\xi_{||}\sum_{n=-N}^{N}\frac{1}{\left[\varepsilon+\frac{\xi_{\perp}^2}{R^2}\left(n-\frac{\Phi}{\Phi_0}\right)^2\right]^{\frac{3}{2}}}. \quad (2)$$

We have concluded that the proposed method of Livanov and Varlamov should be taken as a basis, because the different sequence in the calculations can not give a positive result. For example, we can integrate the expression (1) over momentum in the plane and obtain the following result:

$$\sigma^{xx}(\varepsilon,h,\omega) = \frac{e^2 h}{8s} \sum_{n=0}^{\infty} (n+1) \left\{ \frac{1}{h-i\widetilde{\omega}} \frac{1}{\sqrt{[\varepsilon + h(2n+1)][r + \varepsilon + h(2n+1)]}} + \right.$$

$$+ \frac{1}{h+i\widetilde{\omega}} \frac{1}{\sqrt{[\varepsilon + h(2n+3)][r + \varepsilon + h(2n+3)]}} -$$

$$\left. - \frac{2h}{h^2 + \widetilde{\omega}^2} \frac{1}{\sqrt{[\varepsilon + h(2n+2) - i\widetilde{\omega}][r + \varepsilon + h(2n+2) - i\widetilde{\omega}]}} \right\}. \qquad (3)$$

Now we "roll the plane into a cylinder" by means of summation of the angular quantization levels components (3) up till the last completed level, replacing the summation $n$ of number of equidistant Landau levels $\sum_{n=0}^{\infty}$ in expression (3) with the summation $\sum_{n=-N}^{N}$, where $N$ is a number of subbands filled with electrons, which is determined by the chemical potential and the distance between the levels of quantized angular motion. The reduced magnetic field and the anisotropy parameter respectively are expressed in the following forms:

$$h = \frac{2}{\mu_0} \frac{\xi_{||}^2 \Phi}{R^2 \Phi_0}, \quad r = \frac{4\xi_{\perp}^2}{R^2}. \qquad (4)$$

Thus, we obtain the following expression of paraconductivity in nanotube:

$$\sigma^{||}(h) = \frac{2e^2}{3\mu_0} \frac{\xi_{\perp}^2 \cdot \xi_{||}^2}{R^5} \frac{\Phi}{\Phi_0}. \qquad (5)$$

So, in the final result (5), the paraconductivity dependence on the reduced temperature and the number of subbands in the energy spectrum of the nanotube can not be traced.

Hence, the conclusion is made: during the superconducting nanotubes paraconductivity determination, it is necessary at first to carry out quantization of angular motion, and only after that, the integration over the momentum is possible, because of the noncommutativity of these operations.